# Anchoring Magnetic Fields in Turbulent Molecular Clouds II - from 0.1 to 0.01 parsec

Y. Zhang, Z. Guo, H.H. Wang & H-b Li[*]
Department of Physics, The Chinese University of Hong Kong, Shatin, New Territory, Hong Kong, China.
*hbli@cuhk.edu.hk

## Abstract

We (Li et al. 2009; Paper-I) compared the magnetic field directions inferred from polarimetry data obtained from 100-pc scale inter-cloud media (ICM) and from sub-pc scale molecular cloud cores. The highly correlated result led us to conclude that cloud turbulence must be sub-Alfvenic. Here we extend the study with 0.01-pc cores observed by interferometers. The inferred field directions at this scale significantly deviate from that of the surrounding ICM. An obvious question to ask is whether this high-resolution result contradicts the sub-Alfvenic picture concluded earlier. We performed MHD simulations of a slightly super-critical (magnetic criticality = 2) clouds with Alfvenic Mach number $M_A = 0.63$[1], which can reproduce the Paper-I results, and observed the development towards smaller scales. Interestingly, all subregions hosting cores with $n_{H2} > 10^5$/cc (the typical density observed by interferometers) possess $M_A = 2-3$. Not too surprisingly, these slightly super-Alfvenic cores result in B-field orientation offsets comparable to the interferometer observations. The result suggests that gravity can concentrate (and maybe also contribute to, which takes more study to confirm) turbulent energy and create slightly super-Alfvenic cores out from sub-Alfvenic clouds. The results of our simulations also agree with the observed velocity-scale (Kauffmann et al. 2013), mass-scale (Lombardi et al. 2010) and field strength-density (Li et al. 2015; Crutcher et al. 2010) relations.

## 1. Introduction

The critical role played by turbulence in star formation is solidly established in the past two decades; see, for example, the review by McKee & Ostriker (2007). The review noted that cloud magnetic fields (B-fields) are "not too weak, however" based on the ordered B-fields mapped from three clouds (Li et al. 2006). We (Paper-I) further confirmed this picture of ordered cloud fields by showing the field-direction correlation between 25 cloud cores (1-0.1 pc) and their surrounding inter cloud media (ICM ~ 100 pc). The offsets between the core and ICM fields are shown in Fig. 1 (black). Afterwards, more evidence of dynamically important cloud B-fields have been detected; see our PPVI review (Li et al. 2014).

Recent interferometer (SMA and CARMA) surveys showed that the core-ICM correlation in B-field direction seems not to present within small cores at ~ 0.01 pc (Zhang et al. 2014; Hull et al. 2014). This raises several questions: (1) Is the situation the same for the cores collected in Paper-I? (2) If so, what does the situation imply? Does it mean that the turbulence changes from sub-Alfvenic, the conclusion from Paper-I, to super-Alfvenic at smaller scales? To answer question (1), we have collected data from Zhang et al. (2014) and Hull et al. (2014) (Z/H14 hereafter) for those cores in Paper-I and the result is presented in section 2, Table I and Fig. 1. To answer question (2), we have performed sub-Alfvenic molecular cloud simulations to first reproduce the result from Paper-I (above 0.1 pc) and then observe the circumstance at scales 0.1-0.01 pc. The results of the simulations are summarized in section 3, which are compared with observations in section 4.

---

[1]$M_A \equiv \langle \sigma_V/v_A \rangle$, where $\sigma_V$ and $v_A$ are, respectively, local 3D velocity dispersion and Alfven velocity; $\langle ... \rangle$ means the average within the entire simulated volume (e.g. Burkhart et al. 2009).

## 2. Extending Paper-I to smaller scales

The core-field data in Paper-I are collected by CSO (Dotson et al. 2010) and JCMT (Matthews et al. 2009). The scale is an apparent potential reason for the different field alignment observed by Paper-I and the later interferometer surveys (Z/H14). To test this, we extend the study of Paper-I to smaller scales using data from Z/H14. The field offsets between Z/H14 and the vicinity ICM (adopted from Paper-I) versus core scales is also shown in Fig. 1. The polarization data from Z/H14 for each core is propagated to the mean core field direction using Stokes parameters following Paper-I. A core scale is defined by the 10%-peak intensity contour and estimated by the average of the contour scales in the RA and Dec directions. Indeed, the upper-envelope of Fig. 1 increases as the scale decreases. Below 0.1 pc, there seems no preference for the ICM field direction; the offsets evenly distribute between 0 and 90 degrees.

## 3. MHD simulations

It is suggested in Paper-I and later confirmed by others (e.g., Li et al. 2015b) using numerical simulations that the field alignment seen in Paper-I is possible only if the turbulence is trans- or sub-Alfvenic. The most straightforward observation we should make here on those simulations that already reproduce the result of Paper-I is how the field directions will develop with further fragmentation to compare with Fig. 1. Hence, we performed a set of 3D MHD simulations of sub-Alfvenic clouds using the ZEUS-MP code (Hayes et al. 2006). We observe the change of B-field orientation associated with gas clumps of various scales as seen in Fig. 1. The detailed parameters are described in the following paragraph.

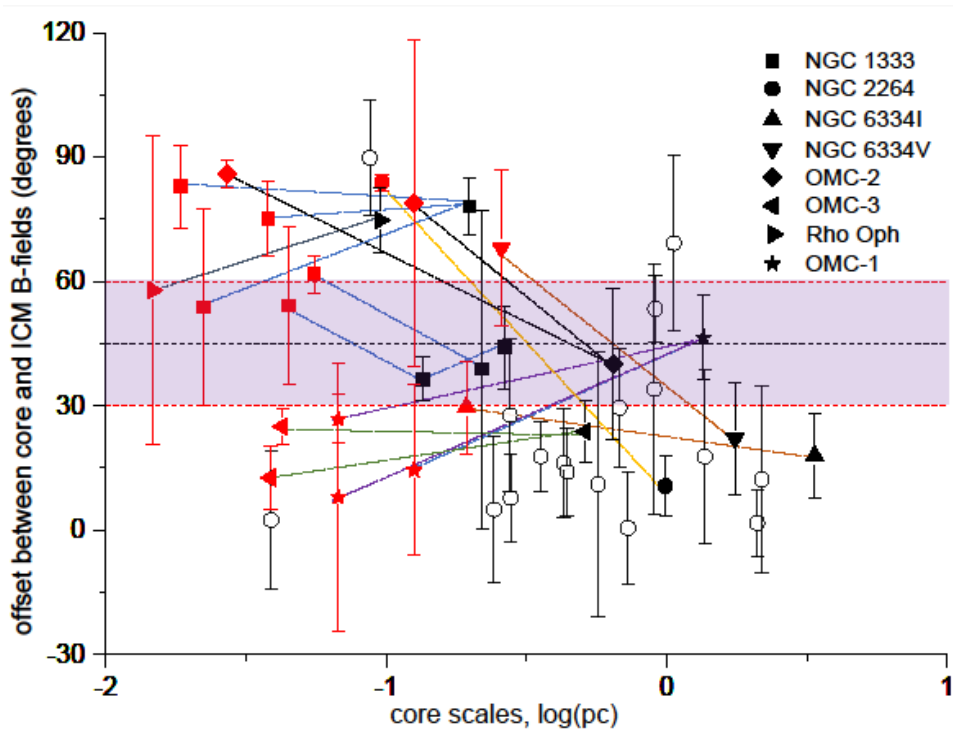

**Fig. 1** Core/ICM B-field offsets against the spatial scales of cores. The black symbols are from the data in Paper-I, and the red symbols are the interferometer data (Z/H14). Data from Paper-I with no interferometer counterparts are shown as hollow circles. The error bars indicate the interquartile ranges (IQRs) of the B-field orientation distributions. Above ~ 0.1 pc, most of the offsets are smaller than 45 degrees and the upper envelope grows with decreasing scale. Below 0.1 pc, there seems no direction preference at all.

3.1 Simulation setup

The simulations are isothermal (10 K) and start with a uniform density ($1.20 \times 10^{-21}$ g/cm$^3$ or 300 $H_2$/cc assuming a mean molecular weight of 2.36) and a uniform B-field (14.4 μG) over a cube with a linear size of 4.8 pc, which is resolved evenly by 960 pixels. The boundary condition is periodic. The ratio of the total mass to the magnetic critical mass ($\Phi/2\pi G^{1/2}$, where $\Phi$ is magnetic flux), a.k.a. magnetic criticality, is 2. The simulations proceed in two stages. First, self-gravity is turned off, and the pure solenoidal turbulence is driven at 2.4 pc until the turbulent energy power spectrum is stable. At this point, the sonic Mach number is 5.7 and the overall Alfvenic Mach number, $M_A \equiv <\sigma_V/v_A>$, as conventionally defined in the literature (e.g. Falceta-Gonçalves et al. 2008, Burkhart et al. 2009), is 0.74. This $M_A$, however, is not observable, as both the Chandrasechar–Fermi and Zeeman methods measure the mean field strength of a targeted volume. To connect simulations with observations, here we define $M_{A\_obs} = \Sigma_V/V_A$, where $\Sigma_V$ and $V_A$ are similar to $\sigma_V$ and $v_A$ but the density weighted 3D velocity dispersion and Alfven velocity based on the density weighted mean field strength of the whole volume; $M_{A\_obs} = 0.92$. The viral parameter, $5\Sigma_V^2 L/6GM$, is 0.51, were $M$ and $L$ are, respectively, the total mass and total volume scale. During the second stage, self-gravity is turned on while the turbulence driving continues for another 1 Myr. At the end, the velocity spectrum remains Kolmogorov-type (Fig. 3) with an overall $M_{A\_obs} = 0.84$ and $M_A = 0.63$. More details about the turbulence driving can be found in the Appendix of Otto, Ji & Li (2017). The Jeans length is always resolved by at least 8 simulation pixels (Truelove et al. 1997). We repeated the simulation three times (Cube 1, 2 and 3; Table II) with different random seeds for turbulence generation.

As our goal is to understand observations, we will mainly use $M_{A\_obs}$ hereafter. Note that, conventionally, only the initial $M_A$ of a simulation is reported in the literature (e.g., Li et al. 2015b, Mocz et al. 2017), but an initial condition is not comparable with observations. The energy redistributes between gravity, turbulence and B-fields as time goes on. Here we report $M_{A\_obs}$ one million years after gravity is turned on, which is in the same order as the typical cloud age.

3.2 Simulation results

**Overall field-density relation**

As the first look of the simulation result, we simply survey the offset between local field orientation and the initial field direction and study how is this offset related to density (Fig. 2, upper panel). For "local" field orientation, we divided the $960^3$-pixel cube evenly into $30^3$-pixel subcubes and calculated the density weighted mean field direction of each subcube. The reason to use the 30-pixel scale for a subcube is that the turbulent energy is artificially dissipated below 20-pixel scale (Fig. 3; which is typical for numerical simulation; see Kritsuk et al. 2011), where the field structures due to turbulence are thus underestimated. The density-offset relation of the $(960/30)^3$ subregions of Cube 1 is shown in Fig. 2 (upper panel), where the upper-envelope of the offset systematically increases with the density. This is the first indication that our simulations might be able to explain Fig. 1, as the upper envelope of Fig. 1 also increases with decreasing scales, which implies increasing density.

Also shown in Fig. 2 (lower panel) is the field strength against density at the pixel scale in Cube 1. The slope of the upper envelop gradually changes from ~ 0.03 at the lower density ($n_{H2} < 10^3$) to ~ 0.67 at the higher density ($n_{H2} > 10^5$).

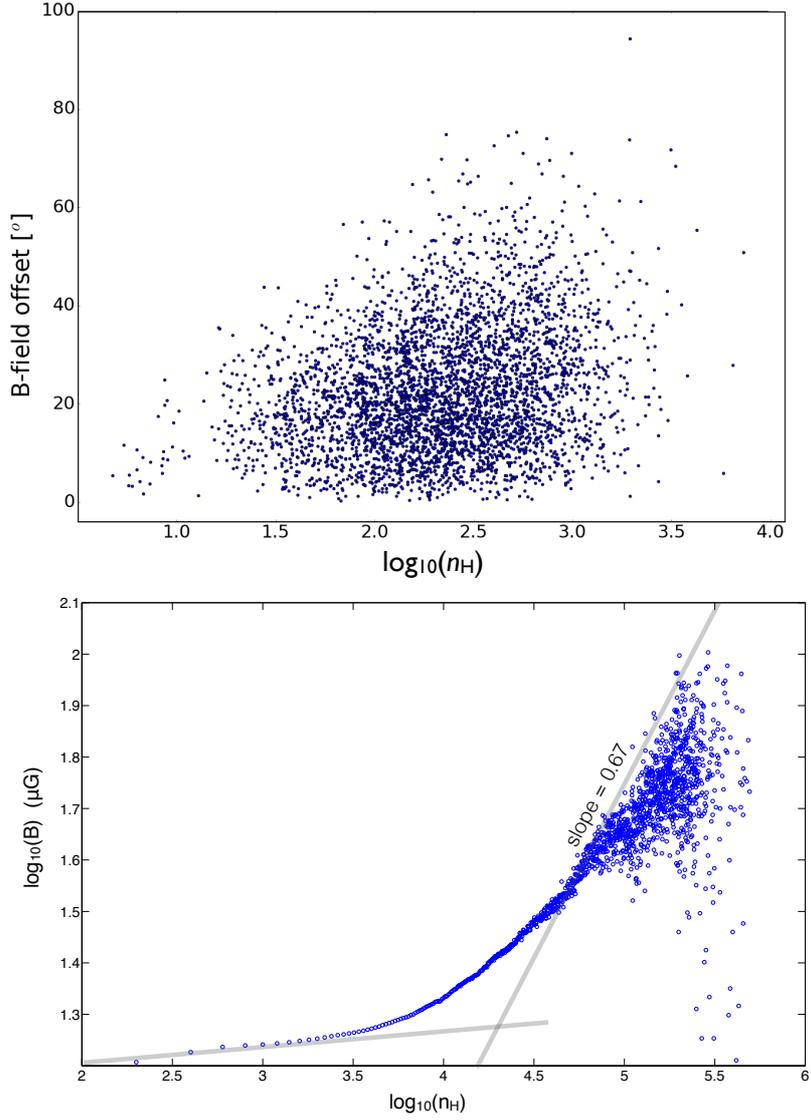

**Fig. 2** B-field versus density in Cube 1. *Upper panel*: Cube 1 is evenly divided into subcubes of $30^3$ pixels, and each data point shows the mean density and the density weighted mean field offset of one subcube. *Lower panel*: The density is binned into groups with an even width of 200 $H_2$/cc. For all the pixels with densities within one bin, their field strengths are averaged. This is a plot of the averaged field strength versus the central density of each bin. The reference slopes of 0.03 and 0.67 are shown as the grey solid lines.

**Field-density relation of high-density cores**

To study how the field can be further deviated in even higher densities, we need to zoom-in into even smaller scales, keeping in mind that once getting smaller than the 20-pixel scale, the field offset from our simulations should be treated as the lower limit because the turbulent energy is artificially dissipated (Fig. 3).

To ensure a density range that covers both single-dish and interferometer data, we first identify cores with $n_{H2} > 10^5$/cc in our simulations; five cores reach this density. For each of these cores and their surrounding regions, contour surfaces of a series of $n_{H2}$, as shown in Table II, are identified. The scale of the volume enclosed by each contour surface is estimated by the cube root of the volume. The mean field direction of each volume is calculated by the density-weighted mean of the field directions at all the pixels within the contour surface. This way, within a core region, density increases with decreasing scale, which imitates the observations in Paper-I and Z/H14.

The result of core field orientations is shown in Table II and Fig. 3, where the 3D offsets are measured from the direction of the uniform field in the initial condition. Fig. 3 already looks inspirational, because of the large offset angles at smaller scales. Above 0.1 pc, which are mainly the scales probed by Paper-I, the angles are all smaller than 45 degrees. Below 0.1 pc, the offset increases but still within 90 degrees. Again, due to the unrealistic turbulence energy dissipation below the 20-pixel scale, the offsets from below 0.1 pc are just lower limits. Mocz et al. (2017) performed similar simulations with a much higher resolution such that artificial energy dissipation only happens at scales much smaller than 0.01 pc and still they found offsets not excessing 90 degree at 0.01 pc for $M_A$ = 1.2 and even 3.5. However, note that their $M_A$ is defined by the "input" kinetic energy and the initial B-field strength; at the snapshot they measured offsets, the $M_A$ should be significantly lower (see section 3.1).

Note that the B-$n_{H2}$ relation "within each core" in Table II is much shallower than a power-law with index 0.67 (Crutcher et al. 2010). The index here (~ 0.3) is closer to the one observed by Li et al. (2015). Indeed, the B-$n_{H2}$ relation of each core describes the condition within "affiliated structures", which is comparable to the observation carried out by Li et al. (2015), but different from Crutcher et al. (2010), where the B-$n_{H2}$ relation is mostly between independent structures. Moreover, the cores in Table II extend to density as low as $n_{H2}$ ~ $10^{3.5}$; if one fits the upper envelope of $n_{H2}$ > $10^{3.5}$ in Fig. 2 (lower panel) with only one power low, the index will be around 0.3. The B-$n_{H2}$ relation in Fig. 2 agrees with Li et al. (2015b) and Mocz et al. (2017) on that the 0.67 index is only for $n_{H2}$ well above $10^4$. In other words, if one power low is fitted to a density range involving $n_{H2}$ < $10^4$, the index has to be significantly less than 0.67. Yet the 0.67 index from Crutcher et al. (2010) covers $n_{H2}$ > 300! This discrepancy between Zeeman measurements and simulations motivated us to revisit the Bayesian analysis of the Zeeman data (Jiang, Li & Fan 2019) and found that the index and the threshold density cannot be reliably derived from the data with large uncertainties in $n_{H2}$.

4. **Discussion**
4.1 Comparison between simulations and observations

Our simulations (Fig. 3) can be compared with the polarimetry observations (Fig. 1) in two different ways. In both cases, we need to get the 2D projections of the field angles. Moreover, if a projected angle is obtuse, the supplementary angle should be adopted to imitate the fact that polarization vectors have the 180-degree ambiguity (headless), and the acute angle is utilized to describe the offset between two polarization vectors (Paper-I; Z/H14).

**The overall trend of the field projections**

We bin the 3D angles into three groups – "above 1 pc", "below 0.1 pc" and "in between". Each 3D angle in Fig. 3 is projected to 135 directions evenly distributed on a sphere centered at the angle. For each bin, all the projected angles are collected (with obtuse angles replaced by their supplementary ones), and their distribution is also shown in Fig. 3. The distributions are comparable to Fig. 1: above 0.1 pc, 90% of the projections are below 45 degrees. Below 0.1 pc scale, even though the 3D offsets here are only the lower limits, the distribution is already close to uniform within 0-90 degrees. In other words, the large field deviations (relative to the local ICM field orientations) observed by interferometers (Fig. 1) do not contradict the idea of an overall sub-Alfvenic cloud ($M_A$ = 0.63).

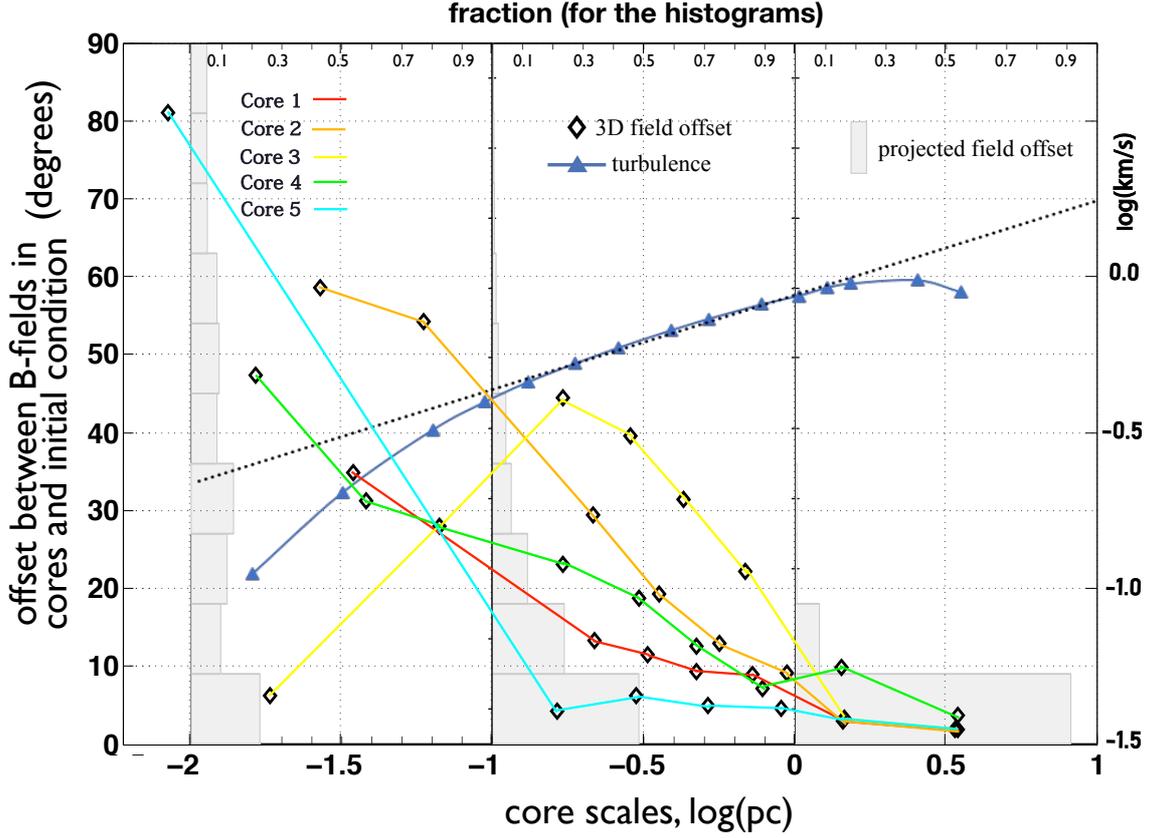

**Fig. 3** The results of MHD simulations. The diamonds show the 3D core field offsets form the initial condition (left axis) against the core sizes (bottom axis; see **Field-density relation of high-density cores** in section 3.2 for the definition of a core size). Each of these 3D offsets is projected along 135 directions evenly distributed over a 4π sr solid angle. The scales are binned into "above 1 pc", "below 0.1 pc" and "in between". The histograms present the distribution of the projected 2D offsets in each bin; the upper axis is the fraction. Note that the 2D histograms are "folded" at 90 degrees to emulate the disability of polarimetry observations on distinguishing supplementary offset angles. Also plotted is the turbulent 3D velocity spectrum (the dark blue line; the velocity is shown in the right axis) of Cube 1, which peaks at 2.4 pc, the turbulence driving scale, and is artificially dissipated below 0.1 pc. The dotted line ($y=0.90\, x^{0.31}$) is a fit to the velocity spectrum between 0.1 and 1.25 pc.

**Reproducing the field directions of affiliated structures**

Besides the overall trend, here we show that the simulation results can also reproduce the observed multi-scale field orientations of individual cloud systems. Orion molecular cloud and Cube 1 are used as an illustration, for that they have more cores (OMC-1, 2 and 3; simulated core 1, 2 and 3) and each OMC core has at least three submillimeter observations at different scales (Table I).

In Fig. 4, OMC-1, 2 and 3 are isolated, respectively, from Fig. 1. The smallest structures are OMC-1 KL NW/SE, OMC-2-FIR3, and OMC-3-MMS5/6 (Table I). For each simulated core, we surveyed all the possible projections. If a projection falls within the error bar of any of the aforementioned five smallest cores and the error bars of its parental structures, we plot this particular projection in Fig. 4. Again, even with the underestimated field structures below the 20-pixel scale, there is no problem for the simulation to reproduce the observed field variation over different scales.

## 4.2 The emergence of super-Alfvenic cores

Our MHD simulations can closely reproduce the field offsets from not only Paper-I but also Z/H14. To further explore the reason of the field offsets and answer question (2), we study the relation between $n_{H2}$ and $M_A$. The subcubes under study are evenly distributed within Cube 1 with the smallest separation between two subcube centers equals to 100 pixels (thus there are overlaps between larger subcubes). The result is given in Fig. 5., where the $M_{A\_obs}$ of core 1, 2 and 3 at ~100- and ~150-pixel scales are also displayed. While the simulation cube is overall sub-Alfvenic, it can produce dense and super-Alfvenic subcubes and cores! Which is most likely due to the mass (and thus kinetic energy) concentration along the B-fields, that results in local kinetic (turbulent) energy enhancement without magnetic field compression. This anisotropic accretion channeled by B-fields is also observed in other simulations and discussed by Otto, Ji & Li (2017). Also, part of the gravitational potential energy may be converted into turbulent energy after the accretion (e.g. Heitsch 2013).

Note that the cores are only slightly super-Alfvenic with $M_{A\_obs} \sim 3$ (Fig. 5; Table II). This level of kinetic energy is not enough to randomize the field orientation. The subcubes with $n_{H2} \sim 10^{3.5}$ in Fig. 2 (upper penal) possess $M_{A\_obs} \sim 3$ (Fig. 5), but their field direction offsets seldom go beyond 90 degrees. It will take an $M_A$ an order of magnitude higher to deviate the field to close to 180 degrees, i.e. to randomize the field, at 0.01-pc scale; see, for example, Fig. 3 of Mocz et al. (2017). Indeed, the 3D offsets shown in Fig. 3 are all within 90 degrees, which is far from random (evenly distributed between 0 and 180 degrees). Note that, however, polarization vectors, which have the ambiguity of 180 degrees, are not able to distinguish between randomness and 3D angles distributed within 0-90 degrees. Thus one should not conclude a random field based merely on Z/H14 data (red in Fig. 1).

## 4.3 Polarization holes

The trend of increasing field offset with decreasing scale (and thus with increasing density) explains the so-called "polarization holes" - the tendency of the decreasing fraction of submm polarization with growing column density (N). In Paper-I, we proposed that polarization holes can occur naturally due to more B-field structures along a line-of-sight with higher N, not necessarily due to the lower grain alignment efficiency in high-density regions as suggested in the literature (e.g. Podoan et al. 2001). The fact that Z/H14 see higher polarization fraction than Paper-I (Tang 2016) supports our proposal. Higher N usually implies a line-of-sight going through a higher density and thus more B-field direction dispersion can be expected (Fig. 2). When these richer field structures are not resolved, they will appear as a lower polarization fraction.

## 5. Summary

We extend the multi-scale study of B-field directions in Paper-I down to 0.01-pc scale using the data from Z/H14. The multi-scale field correlation decreases for scales below 0.1 pc. The directional correlation between 100-pc and 0.1-pc fields (Paper-I) has been understood as that a molecular cloud, as a whole, should be trans- or sub-Alfvenic. Our MHD simulations show that dense cores developing from sub-Alfvenic clouds ($M_{A\_obs} = 0.84$ and $M_A = 0.63$) can be slightly super-Alfvenic ($M_{A\_obs} \sim 3$) to significantly deviate the core fields and thus explain the field offsets observed by interferometers (Z/H14), but not enough to randomize the field directions. The simulation results are also consistent with other observations, e.g., $B \propto n^{0.32\pm0.08}$ for affiliated structures (Li et al. 2015), $B \propto n^{0.67}$ for independent high-density structures (Crutcher et al. 2010), $\sigma_V \propto l^{0.31}$, the linewidth-size relation (Kauffmann et al. 2013) and $M \propto R^{1.54\pm0.12}$, the core mass-size relation for core size R ∈ [0.1 1] pc (Lombardi et al. 2010).

Understanding observations through numerical simulations had become a common practice in modern astronomy. However, we note that $M_A \equiv <\sigma_V/v_A>$ in the initial condition, the conventional description of turbulence/B-field relative importance in a simulation, is not comparable with an observed $M_A$. First, after the turbulence spectrum is stabilized in a simulation, the overall $M_A$ should be significantly lower than the initial condition (section 3.1). This is because part of the kinetic energy is converted to random B-field energy and contributed to the denominator of $M_A$. Second, the observable, $M_{A\_obs} \equiv \Sigma_V/V_A$, the ratio between the density weighted velocity dispersion and the Alfven velocity of the density weighted mean field, will not be affected by random B-fields and thus higher than the $M_A \equiv <\sigma_V/v_A>$ (section 3.1). Third, the localized $M_A$, e.g. $M_A$ of a cloud core, can be significantly different from the overall value of the cloud (Fig. 5).

Finally, polarization offsets distributed over 0-90 degrees (the red data in Fig. 1) can be interpreted in two possible ways: (1) the 3D B-field offsets are simply within 90 degrees; (2) the 3D B-fields offsets are over 0-180 degrees, i.e. random, but polarization offsets cannot go beyond 90 degrees due to the headless nature of polarization orientations. The interpretation (2), however, requires the turbulence to be highly super-Alfvenic (e.g., $M_A$=35 in Fig. 3 of Mocz et al. 2017), but Li et al. (2015b) showed that simulations with $M_A$=10 already conflict with the observations in Paper-I (the black data in Fig. 1).

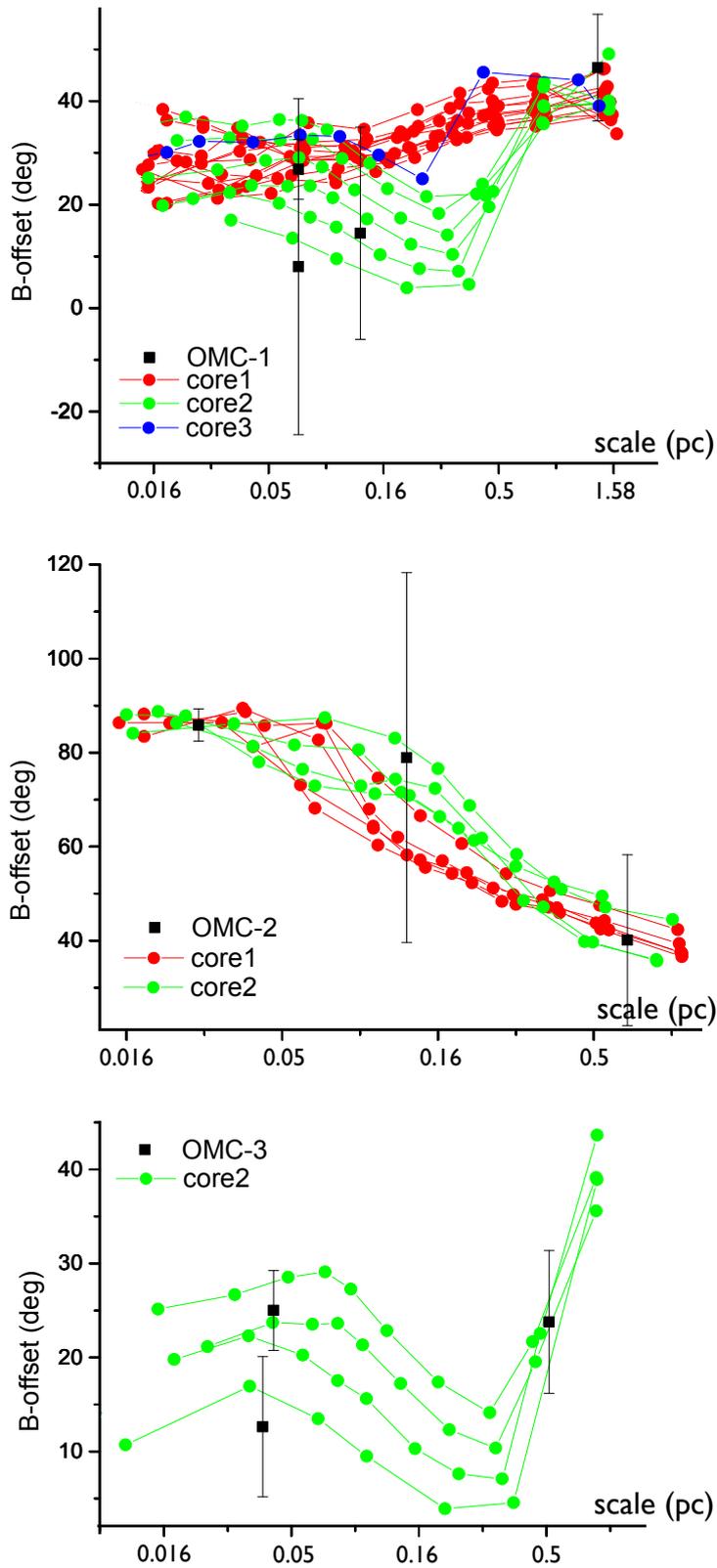

**Fig. 4** The possibility for our simulated sub-Alfvenic cloud (Cube 1) to reproduce OMC field structures (black). Over a scale range of 0.5 pc, the OMC field orientation can change by almost 50 degrees after projected on the plane of sky. Here we survey all the possible projections of core 1-3 in Cube 1 and show that the reproduction of OMC fields is possible. For any affiliated structures (e.g. Orion KL-SE ⊂ Orion KL ⊂ OMC-1), we can always find a simulated core (red, green or blue) with a proper projection to lay the projected field orientations within the error bars from any scale of the affiliated structures.

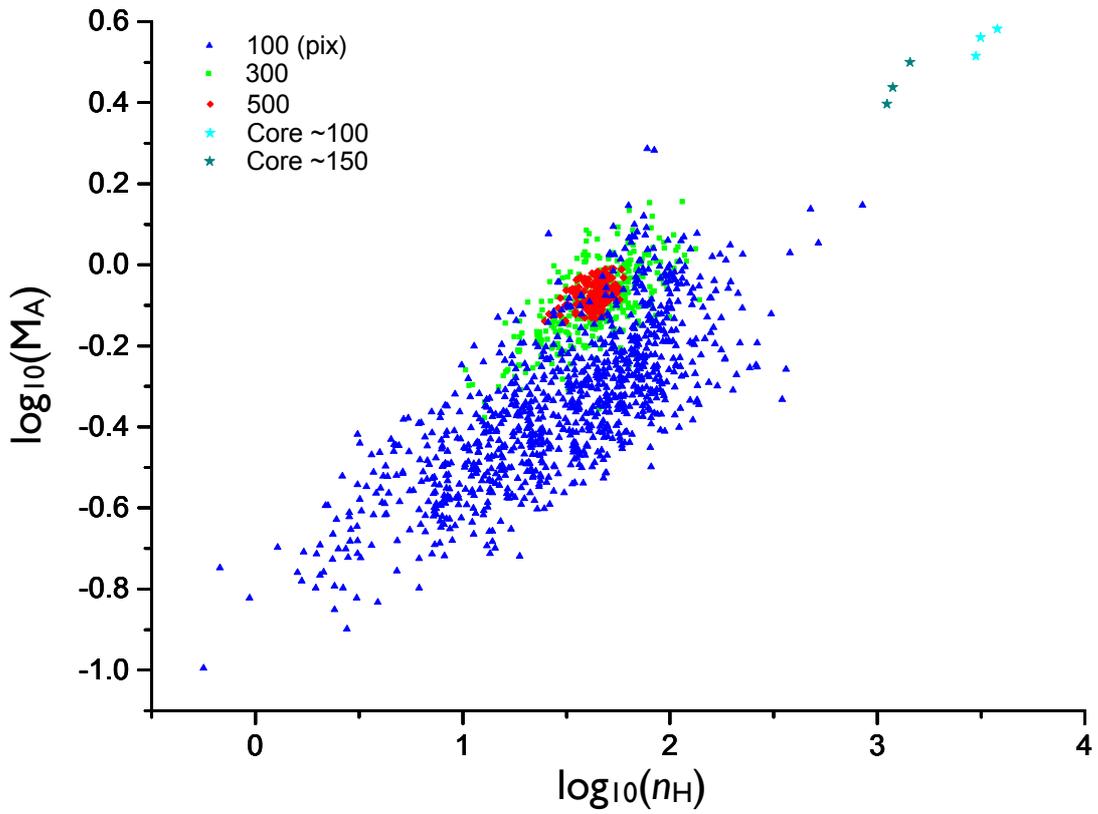

**Fig. 5** Alfvenic Mach number versus density at various scales (100-, 300-, and 500-pixel). The data is from Cube 1. The simulation cube is fully and evenly sampled by subcubes of various sizes, and the $M_{A\_obs}$ and mean density of each subcube are plotted. A positive correlation is shown. The properties of the three dense cores in Cube 1 at 100- and 150-pixel scales are also noted.

Table I. The clouds observed by both Z/H14 and Paper-I

| Object name | Observation | Scale (pc) | Mean B (°) | IQR of B (°) | # of detections |
|---|---|---|---|---|---|
| NGC 1333 | Optical | 300 | -48.5 | 54.3 | 13 |
| NGC 1333 | Hertz CSO | 0.263 | -4.4 | 20.1 | 8 |
| NGC 1333-IRAS 2A | SCUBA JCMT | 0.219 | -9.7 | 76.7 | 22 |
| NGC 1333-IRAS 4 | SCUBA JCMT | 0.197 | 53.4 | 13.7 | 18 |
| SVS 13 | SCUBA JCMT | 0.135 | -12.1 | 10.5 | 43 |
| NGC 1333-IRAS 2A | TADPOL CARMA | 0.0553 | 69.7 | 9.0 | 16 |
| NGC 1333-IRAS 4A | TADPOL CARMA | 0.0379 | 56.4 | 18.1 | 47 |
| NGC 1333-IRAS 4B | TADPOL CARMA | 0.0223 | 77.8 | 47.2 | 18 |
| NGC 1333-IRAS 4B2 | TADPOL CARMA | 0.0184 | 48.6 | 20.0 | 4 |
| SVS 13 | TADPOL CARMA | 0.0450 | 5.6 | 38.1 | 26 |
| | | | | | |
| NGC 2264 | Optical | 300 | -1.8 | 32.4 | 5 |
| NGC 2264 | Hertz CSO | 0.987 | -12.4 | 14.8 | 18 |
| NGC 2264 C1 | SMA | 0.0965 | -85.6 | 3.9 | 12 |
| | | | | | |
| NGC 6334 I | Optical | 300 | -32.4 | 35.2 | 8 |
| NGC 6334 I | Hertz CSO | 3.36 | -50.3 | 20.1 | 32 |
| NGC 6334 I | SMA | 0.193 | -2.2 | 22.3 | 46 |
| | | | | | |
| NGC 6334 V | Optical | 300 | -24.8 | 39.0 | 9 |
| NGC 6334 V | Hertz CSO | 1.748 | -2.8 | 27.1 | 8 |
| NGC 6334 V | SMA | 0.257 | 87.21 | 37.6 | 14 |
| | | | | | |
| OMC-1 | Optical | 300 | -21.4 | 68.25 | 12 |
| OMC-1 | Hertz CSO | 0.206 | -63.1 | 9.6 | 25 |
| Orion KL | TADPOL CARMA | 0.126 | -35.8 | 41.1 | 13 |
| Orion KL NW | TADPOL CARMA | 0.0674 | -48.2 | 11.6 | 34 |
| Orion KL SE | TADPOL CARMA | 0.0674 | -29.3 | 65.0 | 50 |
| | | | | | |
| OMC-2 | Optical | 300 | 70.4 | 53.3 | 93 |
| OMC-2 | Hertz CSO | 0.642 | 30.3 | 36.3 | 17 |
| OMC-2-FIR3 | TADPOL CARMA | 0.02706 | -15.4 | 6.9 | 7 |
| OMC-2-FIR4 | TADPOL CARMA | 0.126 | -18.5 | 78.7 | 29 |
| | | | | | |
| OMC-3 | Optical | 300 | 70.6 | 51.8 | 93 |
| OMC-3 | Hertz CSO | 0.512 | 46.8 | 15.2 | 43 |
| OMC-3-MMS5 | TADPOL CARMA | 0.0386 | 58.0 | 15.0 | 19 |
| OMC-3-MMS6 | TADPOL CARMA | 0.0426 | 45.6 | 8.5 | 27 |
| | | | | | |
| Rho oph | Optical | 300 | -6.1 | 45.7 | 13 |
| Rho oph | Hertz CSO | 0.0694 | 67.6 | 10.0 | 20 |
| VLA 1623 | TADPOL CARMA | 0.0147 | 51.7 | 74.6 | 26 |

Table II. Properties of the cloud cores resulted from our sub-Alfvenic MHD simulations

| | | Boundary $n_{H2}$/cc | Scale(pc) | B offset from initial direction (°) | Mean $n_{H2}$/cc [#] | Mean B (µG)[†] | $M_{A\_obs}$[*] |
|---|---|---|---|---|---|---|---|
| | **Core 1** | | | | | | |
| | | 200000 | 0.035 | 34.9 | 275552 | 48.37 | 2.12 |
| | | 100000 | 0.067 | 28.0 | 151554 | 39.35 | 2.67 |
| | | 20000 | 0.22 | 13.3 | 37255 | 23.30 | 3.83 |
| | | 10000 | 0.33 | 11.5 | 20738 | 19.53 | 3.95 |
| | | 5000 | 0.48 | 9.3 | 11283 | 16.57 | 3.82 |
| | | 2000 | 0.73 | 8.9 | 5338 | 14.60 | 3.16 |
| | **Core 2** | | | | | | |
| **Cube1** | | 200000 | 0.027 | 58.6 | 220300 | 44.74 | 0.92 |
| | | 100000 | 0.060 | 54.3 | 144254 | 33.42 | 2.40 |
| | | 20000 | 0.22 | 29.4 | 32944 | 17.97 | 4.22 |
| | | 10000 | 0.36 | 19.3 | 17696 | 15.80 | 3.78 |
| | | 5000 | 0.57 | 12.9 | 9524 | 15.29 | 3.00 |
| | | 2000 | 0.95 | 9.1 | 4376 | 15.17 | 2.12 |
| | **Core 3** | | | | | | |
| | | 100000 | 0.019 | 6.2 | 111272 | 29.66 | 3.41 |
| | | 20000 | 0.17 | 44.4 | 28798 | 13.35 | 4.20 |
| | | 10000 | 0.29 | 39.6 | 16754 | 12.52 | 4.02 |
| | | 5000 | 0.43 | 31.4 | 9766 | 12.26 | 3.64 |
| | | 2000 | 0.69 | 22.2 | 4620 | 12.80 | 2.74 |
| | **Core 4** | | | | | | |
| | | 200000 | 0.017 | 47.4 | 244180 | 82.13 | 1.46 |
| | | 100000 | 0.039 | 31.3 | 129636 | 50.19 | 1.58 |
| **Cube2** | | 20000 | 0.17 | 23.1 | 31396 | 21.10 | 3.04 |
| | | 10000 | 0.31 | 18.7 | 16564 | 16.17 | 3.53 |
| | | 5000 | 0.48 | 12.5 | 9390 | 14.83 | 3.27 |
| | | 2000 | 0.79 | 7.2 | 4388 | 14.29 | 2.49 |
| | **Core 5** | | | | | | |
| | | 100000 | 0.0085 | 81.1 | 100926 | 28.52 | 0.24 |
| | | 20000 | 0.16 | 4.3 | 29360 | 14.26 | 5.45 |
| **Cube3** | | 10000 | 0.30 | 6.2 | 15720 | 12.83 | 5.21 |
| | | 5000 | 0.52 | 4.9 | 8382 | 12.35 | 4.35 |
| | | 2000 | 0.90 | 4.6 | 4000 | 12.55 | 3.16 |
| | | | | | | | |
| | | 2000; entire Cube 1 | 1.47 | 3.3 | 4172 | 14.26 | |
| | | 200; entire Cube 1 | 3.40 | 1.9 | 860 | 14.26 | |
| | | | | | | | |
| | | 2000; entire Cube 2 | 1.43 | 9.8 | 3948 | 13.97 | |
| | | 200; entire Cube 2 | 3.50 | 3.7 | 780 | 13.97 | |
| | | | | | | | |
| | | 2000; entire Cube 3 | 1.45 | 2.9 | 3740 | 13.92 | |
| | | 200; entire Cube 3 | 3.48 | 1.9 | 802 | 14.12 | |

[*] $M_{A\_obs}$ drops when the scale goes below 0.1 pc for all cores. This is due to the limitation of our MHD simulations, which have to artificially damp the turbulence below 0.1 pc; see Fig. 3.

[#] The density ($n$) - scale ($l$) relation derived from these five cores is $n \propto l^{-0.98\pm0.25}$ and the core mass $\sim (165.1\pm25.3)\times l^{1.54\pm0.12}$ $M_\odot$

[†] The field strength - density relation derived from these five cores is $B \propto n^{0.32\pm0.08}$; see the discussion in section 3.2

# ACKNOWLEDGEMENTS
This research is supported by four grants from the Research Grants Council of the Hong Kong: Early Career Scheme 24300314; General Research Fund 14305717,14600915, and 14304616.

**References**

Blakesley Burkhart, D. Falceta-Goncalves1, G. Kowal1, and A. Lazarian (2009) *The Astrophysical Journal*, *693*, 250.

Crutcher, R. M., Wandelt, B., Heiles, C., Falgarone, E., & Troland, T. H. (2010). Magnetic fields in interstellar clouds from Zeeman observations: inference of total field strengths by Bayesian analysis. *The Astrophysical Journal*, *725*(1), 466.

Dotson, J. L., Vaillancourt, J. E., Kirby, L., Dowell, C. D., Hildebrand, R. H., & Davidson, J. A. (2010). 350 μm Polarimetry from the Caltech Submillimeter Observatory. *The Astrophysical Journal Supplement Series*, *186*(2), 406.

Hayes, J. C., Norman, M. L., Fiedler, R. A., Bordner, J. O., Li, P. S., Clark, S. E., & Mac Low, M. M. (2006). Simulating radiating and magnetized flows in multiple dimensions with ZEUS-MP. *The Astrophysical Journal Supplement Series*, *165*(1), 188.

Heitsch F., 2013, ApJ, 769, 115

Hull, C. L., Plambeck, R. L., Kwon, W., Bower, G. C., Carpenter, J. M., Crutcher, R. M et al. (2014). TADPOL: A 1.3 mm survey of dust polarization in star-forming cores and regions. *The Astrophysical Journal Supplement Series*, *213*(1), 13.

Jiang, H.J., Li, H.-b & Fan, X.D. (2018), under reviewed by ApJ.

Kauffmann, Jens; Pillai, Thushara; Goldsmith, Paul F. (2013) *The Astrophysical Journal*, *779*, 185.

Kritsuk, A. G., Nordlund, Å., Collins, D., Padoan, P., Norman, M. L., Abel, T., ... & Li, P. S. (2011). Comparing numerical methods for isothermal magnetized supersonic turbulence. *The Astrophysical Journal*, 737(1), 13.

Li, H.-b., Griffin, G. S., Krejny, M., Novak, G., Loewenstein, R. F., Newcomb, M. G., ... & Chuss, D. T. (2006). Results of SPARO 2003: mapping magnetic fields in giant molecular clouds. *The Astrophysical Journal*, *648*(1), 340.

Li, H.-b., Dowell, C. D., Goodman, A., Hildebrand, R., & Novak, G. (2009). Anchoring magnetic field in turbulent molecular clouds. *The Astrophysical Journal*, *704*(2), 891.

Li, H.-b., Goodman, A., Sridharan, T. K., Houde, M., Li, Z. Y., Novak, G., & Tang, K. S. (2014). The Link between Magnetic Fields and Cloud/Star Formation. *Protostars and Planets VI*, *1*, 101-123.

Li, H.-b., Yuen, K. H., Otto, F., Leung, P. K. et al. (2015). Self-similar fragmentation regulated by magnetic fields in a region forming massive stars. Nature, 520(7548), 518.

Li, P. S., McKee, C. F., & Klein, R. I. (2015b). Magnetized interstellar molecular clouds–I. Comparison between simulations and Zeeman observations. *Monthly Notices of the Royal Astronomical Society*, *452*(3), 2500-2527.

Lombardi, M.; Alves, J.; Lada, C. J. 2010, A&A,519,7

Matthews, B. C., McPhee, C. A., Fissel, L. M., & Curran, R. L. (2009). The legacy of SCUPOL: 850 imaging polarimetry from 1997 to 2005. *Astrophys. J. Suppl*, *182*, 143-204.

McKee, C. F., & Ostriker, E. C. (2007). Theory of star formation. *Annual Review of Astronomy and Astrophysics*, *45*.

Mocz, P., Burkhart, B., Hernquist, L., Mckee, C., & Springel, V. (2017). Moving mesh simulations of star forming cores in magneto-gravo-turbulence. *The Astrophysical Journal*, 838, 40

Otto, F., Ji, W., & Li, H.-b. (2017). Velocity Anisotropy in Self-gravitating Molecular Clouds. I. Simulation. *The Astrophysical Journal*, *836*(1), 95.

Padoan, P., Goodman, A., Draine, B. T., Juvela, M., Nordlund, Å., & Rögnvaldsson, Ö. E. (2001). Theoretical models of polarized dust emission from protostellar cores. *The Astrophysical Journal*, 559(2), 1005.

Tang, K.S. (2016) Topics of Magnetic Field and Turbulence in Star Formation. A thesis submitted for the Degree of Master of Philosophy in Physics, The Chinese University of Hong Kong (http://sfg.phy.cuhk.edu.hk/group_page/theses/sunny-mphil-thesis.pdf)

Truelove, J. K., Klein, R. I., McKee, C. F., Holliman II, J. H., Howell, L. H., & Greenough, J. A. (1997). The jeans condition: a new constraint on spatial resolution in simulations of isothermal self-gravitational hydrodynamics. *The Astrophysical Journal Letters*, *489*(2), L179.

Zhang, Q., Qiu, K., Girart, J. M., Tang, Y. W., Koch, P. M., Li, Z. Y. et al. (2014). Magnetic fields and massive star formation. *The Astrophysical Journal*, *792*(2), 116.